\documentclass[twocolumn,aps,floatfix]{revtex4}
\usepackage{amssymb}
\usepackage{graphicx}

\begin{document}

\title{Decay of multiply charged vortices at nonzero temperatures}

\author{Tomasz Karpiuk,$^1$ Miros{\l}aw Brewczyk,$^1$ Mariusz Gajda,$^{2,4}$ and 
        Kazimierz Rz\c a\.zewski$\,^{3,4}$}                          

\affiliation{\mbox{$^1$ Wydzia{\l} Fizyki, Uniwersytet w Bia{\l}ymstoku, 
                        ulica Lipowa 41, 15-424 Bia{\l}ystok, Poland}  \\
\mbox{$^2$ Instytut Fizyki PAN, Aleja Lotnik\'ow 32/46, 02-668 Warsaw, Poland} \\
\mbox{$^3$ Centrum Fizyki Teoretycznej PAN, Aleja Lotnik\'ow 32/46, 02-668 Warsaw, 
           Poland}  \\
\mbox{$^4$ Faculty of Mathematics and Sciences UKSW, Warsaw, Poland}  }             

\date{\today}

\begin{abstract}
We study the instability of multiply charged vortices in the presence of thermal atoms and 
find various scenarios of splitting of such vortices. The onset of the decay of a vortex
is always preceded by the increase of a number of thermal (uncondensed) atoms in the system
and manifests itself by the sudden rise of the amplitude of the oscillations of the quadrupole
moment. Our calculations show that the decay time gets shorter when the multiplicity of
a vortex becomes higher.

\end{abstract}

\maketitle

Experiments with atomic Bose-Einstein condensates have already shown their peculiar response
to rotation, manifested through an induced irrotational flow and leading to quantized 
vortices and superfluidity\cite{Cornell,Dalibard,Ketterle,WP2}. It was directly demonstrated 
by using the interferometric technique that the circulation is indeed quantized \cite{Dalibard2} 
and different aspects of dynamics of quantized vortices were investigated 
\cite{Dalibard,Ketterle,Dalibard1,lattice,Ketterle1,Ketterle2}. 
Recently, also the persistent flow of condensed atoms in a toroidal trap was observed \cite{WP2}.

So far studies of quantized vortices were related to the single vortex state, the vortex 
lattices, and the multiply charged vortices. They included the analysis of the process of their 
nucleation as well as the stability conditions and the decay. For example, in Ref. \cite{Dalibard}
the lifetime of the single-vortex state in an axisymmetric trap is investigated at two different
condensate parameters. Although in both cases the uncondensed part of the atomic cloud is almost
undetectable its influence on the lifetime of the vortex is huge. Clearly, reducing the level 
of thermal atoms makes the lifetime of the vortex longer. Surprisingly long vortex lifetimes 
(up to several seconds) were observed for highly ordered triangular vortex lattices containing more
than $100$ vortices \cite{Ketterle}. Multiply charged vortices have additional degree of freedom,
their lifetime is limited by the instability leading to the decay into vortices having smaller
charges. Pioneering work on doubly quantized vortices generated with the help of topological
phase-imprinting technique showed that their lifetime is a monotonic function of the interaction
strength \cite{Ketterle2,Ketterle1}.

Not much experimental work has been devoted to the rotational properties of Bose-Einstein
condensates at finite temperatures. In Ref. \cite{lattice} the crystallization and the decay
of vortex lattices is studied in the presence of thermal atoms. The decay of a lattice was
observed in a nondestructive way just by monitoring the distortion of the rotating condensate
and dramatic dependence on temperature was found. It turned out that the rotational frequency
of the lattice decreases exponentially what has been attributed to the observation that the
thermal cloud was also rotating. In contrast, when the thermal cloud is static 
the decay shows nonexponential behavior \cite{Dalibard}. In Ref. \cite{WP2}, on the other hand, the
influence of the thermal fraction on the stable circulation of condensed atoms was investigated.
It was found that in a trap with ring geometry the persistent flow is possible even with the
condensate fraction as small as $20\%$.

Contrary to the case of zero temperature (for a review see \cite{Fetter} and references
therein) there is only a small number of theoretical papers considering the vortex dynamics 
at the presence of thermal atoms \cite{theory}.

In this Letter, we focus on the evolution of multiply charged vortices generated in a way
described in \cite{WP2} and the role the thermal fraction plays in the splitting process.
Therefore, we use an approach that allows a unified treatment of both condensed and 
thermal atoms. In this method, called the classical fields approximation \cite{przeglad},
the primary object is the complex field which represents all atoms. It is only the measurement
process thereupon the atomic cloud is split in two components: the condensate and the 
thermal atoms. Such a decomposition is inherently built in the detection process since
the detection always takes finite time and is performed under limited spatial resolution. 
In other words, the measurement is a kind of coarse graining procedure, hence eliminating 
some information from the classical field and allowing for the appearance of the thermal 
cloud. On a numerical level, it requires to calculate the time and/or space average of a 
one-particle density matrix built of the classical field. Its dominantly populated eigenmode 
is the condensate wave function, the other modes represent thermal atoms, and its diagonal 
part integrated along the direction of imaging beam is just what is monitored by the CCD 
camera.

Our numerical procedure takes the following steps. First, we find the ground state of the
Bose-Einstein condensate in the toroidal trap made by combining the usual harmonic trap with 
the Gaussian laser beam (as in the experiment of Ref. \cite{WP2}). The harmonic trap frequencies 
are taken as $\omega_z=2\pi \times 25$\,Hz and $\omega_{\perp}=2\pi \times 36$\,Hz.
The blue-detuned laser beam serves as an optical plug repelling atoms from the trap center
and the potential from this beam is given by the Gaussian function $V_0 \exp{[-2 (x^2+y^2)/w_0]}$
(the laser beam propagates in $z$ direction). Here, $V_0$ is the maximum optical potential 
($V_0/h=3600$\,Hz, where $h$ is the Planck constant) and $w_0$ ($w_0=15\,\mu m$) is the waist 
of the beam. Next, we specify the initial condition for the classical 
field by putting a particular amount of energy into the system and allowing the gas to
thermalize. Since the classical field fulfills the time-dependent Gross-Pitaevskii equation 
\cite{przeglad}, we find its evolution and built the one-particle density matrix at each
time. Now, we can do a coarse graining and calculate a fraction of condensed atoms,
the condensate wave function, and the distribution of thermal (uncondensed) atoms.

A recently developed experimental technique, based on a stimulated Raman process with
Laguerre-Gaussian beams \cite{WP1,WP2}, allows to generate multiply charged vortices in 
atomic Bose-Einstein condensates. However, such vortices are unstable against the decay
into a number of vortices with lower charges. It happens because: (i) the energy of a collection
of vortices with lower charges is smaller than the energy of a single multiply quantized
vortex, and (ii) the total circulation is preserved. Now, the question can be raised how the
multiply charged vortex is split into a larger number of vortices. Since the energy
of various configurations of vortices changes, one might expect different scenarios of a
decay.

To investigate the dynamics of the Bose gas after imprinting the phase on it we evolve the 
classical field according to the Gross-Pitaevskii equation. Before we go into details we want
to stress that various scenarios of vortex decay are possible. The way the system follows
depends on how fast the plug is switched off, i.e., on the energy of the gas when the plug is 
off. The shorter removal time ends in higher energy state, therefore allowing the access to higher 
energy configurations. For example, assuming the fivefold charged vortex was imprinted on the
condensate, the considered configurations could be the one with four singly charged vortices
placed on a circle around the vortex located in the center of the trap (called '4+1'
configuration) or the other arrangement with five vortices settled on a circle and no 
vortices in the middle ('5+0' configuration). It can be checked numerically by using the
imaginary time technique in a rotating frame of reference \cite{Castin} that the '4+1' structure 
has higher energy than '5+0' one when it rotates with large enough frequency ($\geq 0.7$ of
radial trap frequency) and the opposite is true for slower rotation (even more, for slower
rotation the obtained configurations are no longer at local energy minima). Therefore, 
depending on the time the plug is turned off the energy related arguments suggest possible 
scenarios of decaying process. Experimentalists could observe the '5+0' structure for slower 
removal of the plug and '4+1' configuration followed by the transition to '5+0' structure when 
the plug is taken off quickly. This scenario gets simplified if even as small as $8\%$
fraction of uncondensed atoms is initially present in the system. In this case the system
goes directly to '5+0' configuration. The first notice on the observation of different vortex 
configurations has been already given \cite{SanFeliu}.

\begin{figure}[htb]
\resizebox{2.9in}{5.5in} {\includegraphics{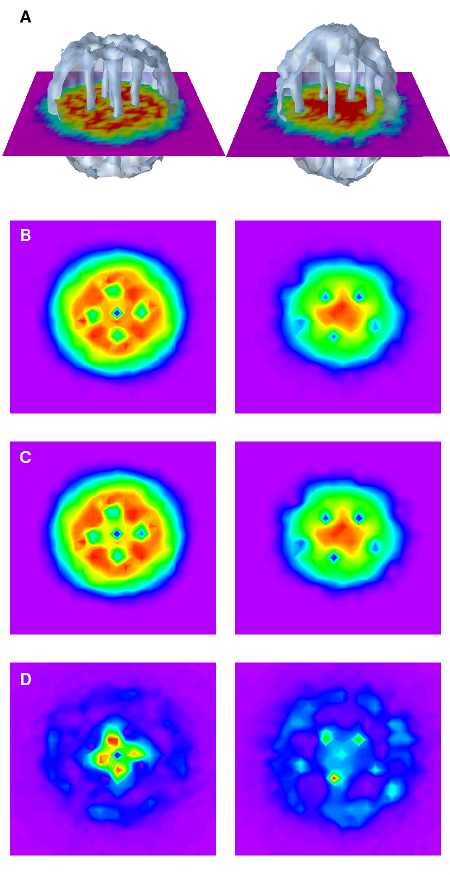}}
\caption{(color online). (A) Isodensity plots, corresponding to the density 
$1.2\times10^{13}$\,cm$^{-3}$, for fast ($30$\,ms, left column) and slow ($200$\,ms, right 
column) removal of the plug showing typical vortex arrays: '4+1' (left frame) and '5+0' 
(right frame) configurations. (B) The $z$-integrated total densities at times and conditions
as in (A). (C),(D) The corresponding condensate and thermal densities, respectively. 
The size of each frame is $55\times 55 \, \mu m$.}
\label{iso1}
\end{figure}

Our numerics confirms just discussed scenarios. Fig. \ref{iso1}A shows the isodensity plots 
at times when the configurations '4+1' (left frame, here the plug is off in $30$\,ms) and
'5+0' (right frame, with the plug taken off in $200$\,ms) are present. The successive rows
display the density as it is imaged by the CCD camera along the direction of axis of symmetry
(B), the $z$-integrated condensate density obtained according to the classical fields
approach (C), and the density of thermal cloud (D). Frames (C) and (D) clearly indicate that
thermal atoms are located in vortices cores. An exception is the vortex placed at the trap
center in '4+1' vortex array which is initially empty. However, immediately it is filled
in with uncondensed atoms it becomes unstable and moves away off the center forming the
'5+0' configuration.

\begin{figure}[htb]
\resizebox{2.9in}{1.6in} {\includegraphics{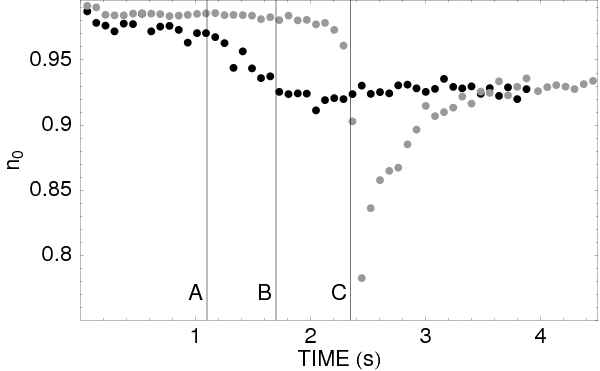}}
\caption{Condensate fraction as a function of time for fast ($30$\,ms, black dots)
and slow ($200$\,ms, dark dots) removal of the plug. Vertical lines indicate:
the times when the '4+1' (A) and '5+0' (B) configurations appear in the case of
fast change of the trapping potential and the time when the '5+0' (C) structure is
visible for slow removal of the plug.}
\label{frac}
\end{figure}

\begin{figure}[htb]
\resizebox{3.4in}{2.0in} {\includegraphics{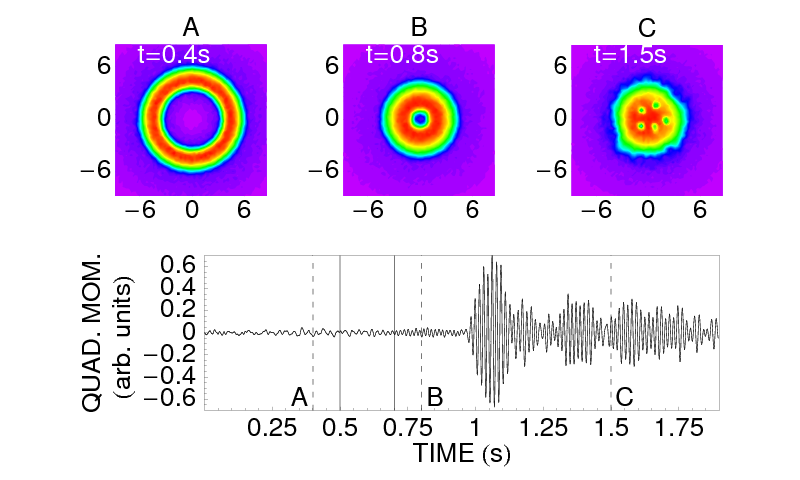}}
\caption{(color online). Decay of a fivefold charged vortex. The upper panel shows the
density at various times (from left to right): while the plug is on (A), after the plug
is off (which takes $200$\,ms) but before the vortex is split into the singly charged vortices (B), 
and when the vortex is clearly split into five vortices (C). In the lower frame the quadrupole
moment is plotted as a function of time showing clearly the onset ($0.3$\,s after the plug
is taken off) of the decay process.  }
\label{decay}
\end{figure}

Details of the dynamics of fivefold charged vortex are given in Figs. \ref{frac} and
\ref{decay}. Fig. \ref{frac} shows the fraction of uncondensed atoms appearing in the system
in the case of fast ($30$\,ms) and slow ($200$\,ms) change of the trapping potential.
Fast removal of the inner plug means stronger disturbance of the gas, hence more effective
production of thermal atoms. As it was already discussed in Ref. \cite{KG}, the level of the 
thermal noise strongly influences the lifetime of the vortex. The larger number of thermal 
atoms the shorter lifetime of the vortex. The fivefold vortex decays approximately $1$\,s
after the plug is taken out (vertical line (A) in Fig. \ref{frac}) and since the fast 
removal of the plug ends in a higher energy state, the higher energy vortex configurations 
are accessible, in our case it is '4+1' configuration. Afterwords, the singly charged vortices
settled on a circle increase their separation from the vortex located at the center of the 
trap and simultaneously the frequency of the 4-vortex array rotation decreases. In other words,
the 4-vortex array looses its energy and the lost energy goes to the thermal atoms. At
some time (line (B) in Fig. \ref{frac}) the frequency the 4-vortex array rotates approaches
the critical frequency ($\approx 0.7$ of radial trap frequency) and the system jumps to the
'5+0' vortex configuration increasing further the fraction of thermal atoms. Finally, both
for fast and slow removal of the optical plug the system ends with '5+0' configuration
rotating with approximately the same frequency. The energy difference between the initially
imprinted the fivefold vortex and the final '5+0' vortex array is transformed to the
thermal modes as can be verified by noticing that the thermal fraction in both cases
remains approximately on the same level (see Fig. \ref{frac}).

\begin{figure}[htb]
\resizebox{3.1in}{2.3in} {\includegraphics{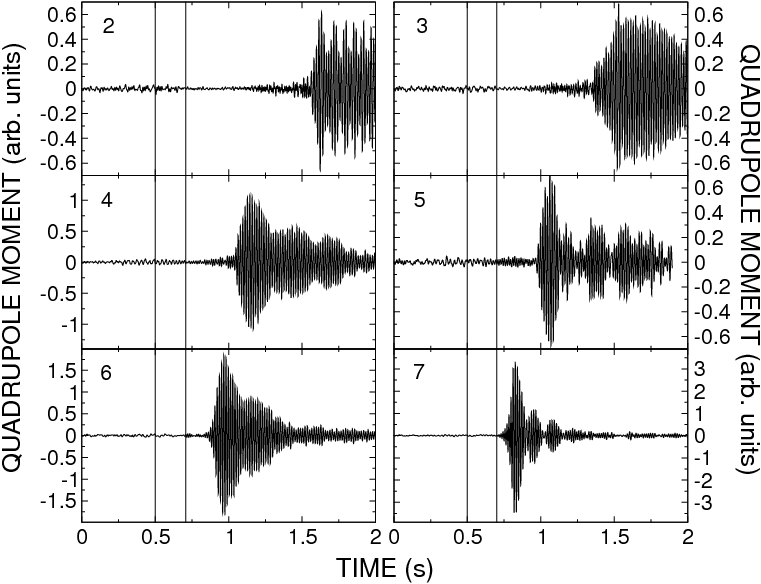}}
\caption{Quadrupole moment as a function of time for differently charged vortices (from 
doubly charged to sevenfold charged as indicated in the left upper corner of each frame). 
Vertical lines show the beginning and the end (delayed by $200$\,ms) of removal of the
plug. Note that, in fact, the dramatic change in the behavior of the quadrupole moment 
determines the onset of the decay of a vortex.}
\label{quad}
\end{figure}

Our calculations show that the scenario of decay of a multiple vortex 
depends strongly on whether initially the Bose gas has a distinguished thermal fraction.
In fact, the way the system behaves determines the upper limit for a number of uncondensed 
atoms. We found that already for the condensate fraction $92\%$ the '4+1' configuration 
is not present irrespective of the duration the optical plug is taken off.
Fig. \ref{decay} shows the evolution of the system for slow removal of the plug when the 
initial condensate fraction equals $n_0=0.92$. In addition, we plot the time dependence of the 
quadrupole moment defined as $\int (x^2-y^2) |\psi(\mathbf{r})|^2 d^{\,3}r$, where 
$\psi(\mathbf{r})$ 
is the classical field. It turns out that the behavior of the quadrupole moment exhibits
clearly the time the multiple vortex begins to split. The decay of the vortex is accompanied
by the sudden increase of the amplitude of the quadrupole moment and allows unambiguously
to determine the lifetime of the vortex for any vortex charge (see Fig. \ref{quad}).

\begin{figure}[htb]
\resizebox{2.9in}{1.8in} {\includegraphics{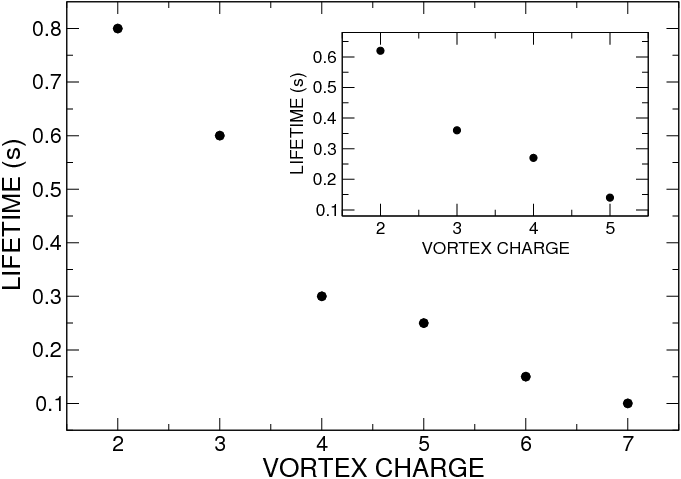}}
\caption{Decay time as a function of the charge of a vortex for a given initial fraction of 
condensed atoms ($n_0=0.92$) for slow ($200$\,ms) and fast ($30$\,ms, Inset) removal of the 
plug.}
\label{time}
\end{figure}

\begin{figure}[htb]
\resizebox{2.9in}{1.6in} {\includegraphics{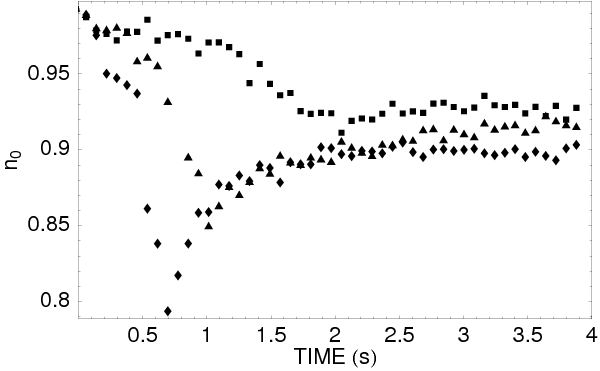}}
\caption{Condensate fraction as a function of time for fast ($30$\,ms) removal of the 
optical plug. Squares, triangles, and diamonds correspond to the charge of the imprinted 
vortex equal to $5$, $6$, and $7$, respectively.  }
\label{frac567}
\end{figure}

Finally, in Fig. \ref{time} we plot the decay time of the multiple charged vortex as a
function of its charge, assuming it is the time that elapsed since the plug beam was off.
Here, the initial condensate fraction equals $n_0=0.92$. The main frame
corresponds to slow change of the trapping potential ($200$\,ms), whereas the inset shows
the case of fast change ($30$\,ms). First observation is that faster removal of the plug
leads to shorter decay times. This feature can be understood based on Fig. \ref{frac}.
Faster removal produces more thermal atoms and hence the destabilization of the vortex
happens in a shorter time (see also Ref. \cite{KG}). The same arguments help to understand
the dependence of the lifetime on the charge of the vortex. As it is displayed in
Fig. \ref{frac567} the production of uncondensed atoms is enhanced when the topological
charge of the vortex is bigger.

In summary, we have studied the decay process of multiply charged vortices. We show that
various scenarios of vortex splitting are possible depending on the level of uncondensed 
atoms appearing in the system as a result of a change of the trapping potential. Initially, 
an amount of uncondensed atoms is determined by how fast the plug supporting the toroidal 
trap is removed. Faster removal produces more thermal atoms and leads to quicker decay,
however, afterwords possibly displaying different vortex configurations. Finally, both fast 
and slow removal results in the same vortex configuration. At the time the vortex decays 
the large oscillations of the quadrupole moment appear. The decay time depends also on the 
multiplicity of the vortex and for even a tiny uncondensed fraction gets shorter for higher 
multiplicity.

\acknowledgments
The authors acknowledge support by the Polish KBN Grant No. 1 P03B 051 30
and by Polish Government research funds for 2006-2009.
Some of the results have been obtained using computers at the Interdisciplinary
Centre for Mathematical and Computational Modeling of Warsaw University.


\begin{thebibliography}{99}

\bibitem{Cornell}
M.R. Matthews, B.P. Anderson, P.C. Haljan, D.S. Hall, C.E. Wieman, and
E.A. Cornell, Phys. Rev. Lett. {\bf 83}, 2498 (1999).
\bibitem{Dalibard}
K.W. Madison, F. Chevy, W. Wohlleben, and J. Dalibard, Phys. Rev. Lett.
{\bf 84}, 806 (2000).
\bibitem{Ketterle}
J.R. Abo-Shaeer, C. Raman, J.M. Vogels, and W. Ketterle, Science {\bf 292},
476 (2001).
\bibitem{WP2}  C. Ryu, M.F. Andersen, P. Clad\'e, V. Natarajan, K. Helmerson, 
and W.D. Phillips, Phys. Rev. Lett. {\bf 99}, 260401 (2007).
\bibitem{Dalibard2} F. Chevy, K.W. Madison, V. Bretin, and J. Dalibard,
Phys. Rev. A \textbf{64}, 031601(R) (2001).


\bibitem{Dalibard1} F. Chevy, K.W. Madison, and J. Dalibard, Phys. Rev. Lett.
{\bf 85}, 2223 (2000).
\bibitem{lattice} J.R. Abo-Shaeer, C. Raman, and W. Ketterle,
Phys. Rev. Lett. {\bf 88}, 070409 (2002).


\bibitem{Ketterle2} Y. Shin, M. Saba, M. Vengalattore, T.A. Pasquini,
C. Sanner, A.E. Leanhardt, M. Prentiss, D.E. Pritchard, and W. Ketterle,
Phys. Rev. Lett. {\bf 93}, 160406 (2004).
\bibitem{Ketterle1} A.E. Leanhardt, A. G\"orlitz, A.P. Chikkatur,
D. Kielpinski, Y. Shin, D.E. Pritchard, and W. Ketterle,
Phys. Rev. Lett. {\bf 89}, 190403 (2002).


\bibitem{Fetter} A.L. Fetter and A. Svidzinsky, J. Phys.: Condens. Matter {\bf 13},
R135 (2001).
\bibitem{theory} 
P.O. Fedichev and G.V. Shlyapnikov, Phys. Rev. A {\bf 60}, R1779 (1999);
C.W. Gardiner, J.R. Anglin, and T.I.A. Fudge, J. Phys. B {\bf 35}, 1555 (2002);
H. Schmidt, K. G\'oral, F. Floegel, M. Gajda, and K. Rz\c a\.zewski,
J. Opt. B {\bf 5}, S96 (2003);
C. Lobo, A. Sinatra, and Y. Castin, Phys. Rev. Lett. {\bf 92}, 020403 (2004);
A.S. Bradley, C.W. Gardiner, and M.J. Davis, Phys. Rev. A {\bf 77}, 033616 (2008).




\bibitem{przeglad} M. Brewczyk, M. Gajda, and K. Rz\c a\.zewski, J. Phys. B {\bf 40}, 
R1 (2007).



\bibitem{WP1} M.F. Andersen, C. Ryu, P. Clad\'e, V. Natarajan, A. Vaziri,
K. Helmerson, and W.D. Phillips, Phys. Rev. Lett. {\bf 97}, 170406 (2006).



\bibitem{Castin} Y. Castin and R. Dum, Eur. Phys. J. D {\bf 7}, 399 (1999).



\bibitem{SanFeliu} W.D. Phillips, {\it Bose-Einstein Condensation 2007} conference
(Sant Feliu, Spain, 2007).


\bibitem{KG} K. Gawryluk, M. Brewczyk, and K. Rz\c a\.zewski, J. Phys. B {\bf 39}, 
L225 (2006).









\end{thebibliography}
\end{document}